\documentclass[12pt,preprint]{aastex}
\usepackage{graphics,graphicx}
\def \s{~\rm{s}}
\def \km{~\rm{km}}

\def \AU{~\rm{AU}}
\def \erg{~\rm{erg}}

\def \yr{~\rm{yr}}

\def \lesssim{\mathrel{<\kern-1.0em\lower0.9ex\hbox{$\sim$}}}
\def \gtrsim{\mathrel{>\kern-1.0em\lower0.9ex\hbox{$\sim$}}}

\begin{document}

\title{X-rays from the Mira AB Binary System}

\author{Joel H. Kastner\altaffilmark{1} and Noam Soker\altaffilmark{2}}

\altaffiltext{1}{Chester F. Carlson Center for Imaging
Science, Rochester Institute of Technology, 54 Lomb Memorial
Dr., Rochester, NY 14623; jhk@cis.rit.edu}
\altaffiltext{2}{Department of Physics, Technion-Israel
Institute of Technology, Haifa 32000, Israel;
soker@physics.technion.ac.il} 

\begin{abstract}
We present the results of XMM-Newton X-ray observations of the
Mira AB binary system, which consists of a pulsating,
asymptotic giant branch primary and nearby ($\sim0.6''$
separation) secondary of uncertain nature. The EPIC
CCD (MOS and pn) X-ray spectra of Mira AB are relatively soft,
peaking at $\sim1$ keV, with only very weak emission at
energies $> 3$ keV; lines of Ne {\sc ix}, Ne {\sc x}, and O
{\sc viii} are apparent. Spectral modeling indicates a
characteristic temperature $T_X \sim 10^7$ K and intrinsic
luminosity $L_X \sim 5\times10^{29}$ erg s$^{-1}$, and
suggests enhanced abundances of O and, possibly, Ne and Si
in the X-ray-emitting plasma. Overall, the X-ray spectrum
and luminosity of the Mira AB system more closely resemble
those of late-type, pre-main sequence stars or late-type,
magnetically active main sequence stars than those of
accreting white dwarfs. We conclude 
that Mira B is most likely a late-type, magnetically active,
main-sequence dwarf, and that X-rays from the Mira AB system
arise either from magnetospheric accretion of wind material
from Mira A onto Mira B, or from coronal activity associated
with Mira B itself, as a consequence of accretion-driven
spin-up.  One (or both) of these mechanisms
also could be responsible for the recently discovered, point-like
X-ray sources within planetary nebulae.
\end{abstract}


\section{Introduction}

The origin and nature of X-ray emission from highly
evolved, post-main sequence stars remains uncertain. In
late-type main-sequence stars, X-ray emission is generally
assumed to trace surface magnetic activity ultimately
derived from stellar magnetic dynamos.
The X-ray detection of several first-ascent red giant
stars by ROSAT (e.g., Schr\"oder, H\"unsch, \& Schmitt 1998;
H\"unsch et al.\ 2003) and, more
recently, Chandra (H\"unch 2001) therefore suggests that
low-mass, post-main sequence stars can be magnetically
active (although it is also possible that the X-rays originate
from active, main-sequence companion stars). On the other
hand, it appears that --- despite maser 
measurements of large local magnetic fields in asymptotic giant
branch (AGB) star winds (e.g.,
Vlemmings, Diamond, \& van Langevelde 2002) --- single
AGB stars are, at best, only weak X-ray sources
(Kastner \& Soker 2004). This indicates that the surface magnetic
fields of AGB stars are locally rather than globally strong (Soker \&
Kastner 2003, hereafter Paper I, and references therein).

The lack of X-ray emission from AGB stars is intriguing and
puzzling, given that several planetary nebulae, as well as
certain ionized, bipolar 
nebulae associated with symbiotic stars, are now known to
harbor point-like X-ray emission at their cores (Chu et al.\
2001; Guerrero et al.\ 2001; Kellogg et al.\ 2001; Kastner
et al.\ 2003). These X-ray sources might be ascribed to PN
central stars whose magnetic fields are sufficiently
powerful to launch and/or collimate their mass
outflows (Blackman et al.\ 2001a). Alternatively, the X-rays
from these systems may 
originate with binary systems, in which the companion is
either accreting material from the mass-losing primary or is
itself magnetically active (Guerrero et al.\ 2001; Soker \&
Kastner 2002, hereafter SK02). In such systems the presence of an accreting
companion, as opposed to magnetic fields on the primary,
would explain the collimation of outflows (Soker \&
Rappaport 2000 and references therein).
Blackman et al.\ (2001b) have proposed that both mechanisms
might operate in certain systems that display multipolar
symmetry.

The Mira (omicron Ceti) system provides a nearby
($D\sim128$ pc) example of a binary
system consisting of a mass-losing AGB star and nearby
($0.6''$ separation) companion (e.g., Karovska et al.\
1997). The nature of the companion is uncertain, as its
optical through UV spectrum appears to be dominated by
emission from an accretion disk that presumably is
accumulated from Mira A's wind (Reimers \& Cassatella 1985;
Bochanski \& Sion 2001; Wood, Karovska \& Raymond 2002).
The Mira AB system was detected as a weak
X-ray source by the Einstein and ROSAT X-ray observatories
(Jura \& Helfand 1984, hereafter JH84; Karovska et al.\ 1996)
but the origin of this X-ray emission is unknown, given the
uncertainty concerning the nature of Mira B. If the
X-rays arise from accretion onto Mira B, then the modest
X-ray luminosity of the system appears to rule out the
possibility that Mira B is a white dwarf (JH84). However,
the early Einstein observations of the system lacked the
sensitivity and X-ray spectral response to provide a
definitive test, in this regard.

In Paper I, we analyzed the archival ROSAT data obtained for
Mira AB in the context of the possibility that the X-rays
arise in magnetic activity on the AGB star (Mira A). While
fits of coronal plasma models to the ROSAT
Position Sensitive Proportional Counter spectrum were
consistent with such a hypothesis, these results left open
alternative possibilities; for example, the X-ray 
emission might arise from magnetic activity on Mira B, or from an
accretion disk around this companion star. Furthermore,
ROSAT lacked hard ($>2.5$ keV) X-ray sensitivity.

Here, we report XMM-Newton observations of Mira AB. The
sensitivity, energy coverage, and spectral resolution of
XMM-Newton far surpass those of Einstein and ROSAT. Hence,
these observations allow us to further constrain the various
models for the Mira AB system. 

\section{Observations and Data Reduction}

XMM-Newton (Jansen et al.\ 2001) observed the Mira system
for 12.22 ks on
2003 July 23. The integration times with the European Photon
Imaging Camera (EPIC) MOS and EPIC pn CCD detector systems
were 11.97 ks and 10.34 ks, respectively. The spectral
resolution of these CCD systems range from
$\sim50$ eV to $\sim150$ eV over the energy range $0.1-10$ keV. The thick
blocking filter was used to suppress optical photons from
the Mira system. Standard X-ray event pipeline processing
was performed by the XMM-Newton Science Center using version
5.4.1 of the XMM-Newton Science Analysis System
(SAS\footnote{http://xmm.vilspa.esa.es/sas/}).

The observation resulted in the detection by EPIC of
$\sim60$ sources in a $\sim25'\times25'$ field centered near
the position of Mira A (= HD 14386; SIMBAD position
$\alpha_{J2000} =$ 02:19:20.7927, $\delta_{J2000} =$
-02:58:39.513). These detections include one source (at
$\alpha_{J2000} =$ 02:19:20.81, $\delta_{J2000} =$
-02:58:41.1) that is consistent with the coordinates of the
Mira AB system, given the astrometric precision of
XMM-Newton (rms pointing uncertainty $>3''$ with a median
pointing error of
$\sim1''$\footnote{http://xmm.vilspa.esa.es/docs/documents/CAL-TN-0018-2-1.pdf}).
We are unable to 
establish whether this emission arises from Mira A or Mira
B, as the latter is found only $0.58''$ from Mira A at
position angle $108^\circ$ (Karovska et al.\ 1997).

We used SAS and the Interactive Data Language to extract
spectra and light curves of 
this source from the MOS 1, MOS 2, and pn event data within
circular regions of radius 20$''$. Background
was determined from an annulus with inner and outer radii of
20$''$ and 40$''$, respectively (the X-ray count rates in
these background regions were comparable to those obtained
for regions farther off source). The
resulting, background-subtracted count rates for the Mira AB
system were 0.027$\pm0.003$ s$^{-1}$ for MOS 1 and MOS 2
(combined) and 0.031$\pm0.003$ s$^{-1}$ for pn.


\section{Results}

\subsection{XMM/EPIC X-ray Spectra}

In Fig.\ 1 we display the combined EPIC (MOS 1, MOS2, and
pn) counts spectrum of the Mira system. The spectrum peaks
at $\sim0.9$ keV (top panel), and emission lines of Ne~{\sc ix},
Ne~{\sc x}, and O~ {\sc viii} appear to be present (bottom
panel). There is little emission at energies $> 3$ keV.

We fit the EPIC spectra of the Mira system using XSPEC
version 11.2 (Arnaud 1996). SAS was used to construct
response matrices and effective area curves for the specific
source spectral extraction regions. Motivated by the
apparent emission lines of Ne and O in the merged EPIC
spectrum, we used a variable-abundance MEKAL model (Kaastra
et al.\ 1996) to fit the spectra (Fig.\ 2). In our spectral
fits, the intervening absorbing column ($N_H$) and X-ray
emission temperature ($T_X$) were taken as free parameters
as the abundances of individual elements were systematically
varied. This procedure was applied during
simultaneous fits to the MOS 1, MOS2, and pn spectra, and the
results were confirmed via independent fits to
each of these three spectra. The best-fit model has an
oxygen abundance that is 
enhanced by a factor $23\pm6$ relative to solar. The fit
results further suggest that the Ne and Si abundances may be
somewhat enhanced ($\sim4$ and $\sim2.5$ times solar,
respectively, with large uncertainties), while the abundance of
Fe is solar (to within the fit uncertainties). These results
for the abundances in the X-ray-emitting gas
are somewhat tentative, however; the best-fit,
variable-abundance model yields $\chi^2 = 0.81$, whereas a
model with all elemental abundances fixed at solar (for
which the best-fit values of $N_H$ and $kT_X$ are similar to
those of the variable-abundance model) yields $\chi^2 = 1.27$. 

From the variable-abundance model, we find best-fit values
for the intervening absorbing column 
of $N_H \approx 4.5\times10^{21}$ cm$^{-2}$ and X-ray emission
temperature of $T_X \approx 10^7$ K, with formal uncertainties of
$\sim20$\%. These results are not very sensitive to the
precise values of the abundances in the model. The best-fit model flux
is $6.4\times10^{-14}$ erg cm$^{-2}$ s$^{-1}$ (0.3--3.0
keV), and the intrinsic (unabsorbed) luminosity derived from
the model is $L_X \approx 5\times10^{29}$ erg s$^{-1}$. The
values for $T_X$ and source flux 
derived from the EPIC data are very similar to those derived
from model fitting of the ROSAT data for the Mira system
(Paper I), while the values of $N_H$ and $L_X$
obtained here are somewhat larger
than those obtained from the ROSAT data. 
Fixing $N_H = 2\times10^{21}$ cm$^{-2}$ (the value
determined from the ROSAT PSPC data) does not change
appreciably the EPIC fit results for $T_X$, but would imply
that the overabundance of O is much more modest ($\sim1.5$ times solar) and,
in addition, that Fe may be depleted ($\sim0.2$
times solar) in the X-ray-emitting plasma. 

We also attempted to fit the EPIC CCD spectra with
two-component thermal plasma models wherein $N_H$ is fixed
at $2\times10^{20}$ cm$^{-2}$, the
value derived from UV spectral modeling (Wood et al.\ 2002). We find
that the fit in this case 
essentially reverts to an isothermal model --- that
is, the ``cool'' component contributes negligibly to
the emission --- but with an unrealistically high plasma
temperature for the ``hot'' component. The result is a
very poor fit, particularly in the 1 keV region where
the Ne lines appear. We conclude that the X-ray-derived value of
$N_H \sim (2-5)\times10^{21}$ cm$^{-2}$ is relatively
robust.

\subsection{XMM/EPIC X-ray Light Curve}

In Fig.\ 3, we display the combined EPIC (MOS 1 + MOS 2 +
pn) light curve of the Mira system. The EPIC count rate is
observed to rise rather abruptly, by a factor $\sim2$,
within the first 3 ks of the observation. The X-ray flux
then more or less steadily declines, such that by the end of
the $\sim10$ ks period during which all 3 CCD cameras were
active, the count rate had returned approximately to a value
at or below that at observation start. The shape of the
X-ray light curve is thus suggestive of a magnetic flare or
enhanced accretion rate event,
although the time interval appears too short to ascertain the
quiescent X-ray count rate and, hence, the characteristic
flare timescale and amplitude.

\section{Discussion}

Before considering the most likely sources of the X-ray
emission from the Mira AB system, we first mention two
mechanisms that are unlikely to contribute to this
emission. One potential source of X-ray emission is that of
collisions between winds from components A and B. It is
difficult to estimate the likely X-ray luminosity due to
such wind shocks, as the mass loss rate ($\dot M_B$) and
wind speed ($v_B$) of
Mira B appear to be strongly variable (Wood et al.\
2002). Recent HST UV observations indicate 
that $\dot M_B \simeq 5 \times 10^{-13} M_\odot \yr^{-1}$ and
$v_B = 250 \km \s^{-1}$ whereas earlier IUE observations
suggest $\dot M_B \simeq 10^{-11} M_\odot \yr^{-1}$ and
$v_B = 400 \km \s^{-1}$ (see Fig.\ 12 in Wood et al.\ 2002). Nevertheless,
adopting these values as representative of high and low
states of mass loss from Mira B, we estimate that the shocked wind X-ray
luminosity is only
$L_x({\rm wind})=5\times 10^{27}-2.5 \times 10^{29} \erg
\s^{-1}$, where these estimates are obtained under the
assumption that the half of the shocked wind
that is expelled toward Mira A 
emits X-rays at $100\%$ efficiency. Therefore, as the conversion of
wind to radiant energy is probably quite inefficient, the
wind shock
mechanism is unlikely to account for the measured X-ray luminosity
of Mira AB (\S 3.1).  

In addition, the Bondi-Hoyle accretion radius (measured from
the center of Mira B) is $R_{\rm acc} = 2 G M_B/v_r^2= 18
\AU$ for reasonable values of the relevant parameters.  This
is much larger than the stagnation distance of the two winds
(measured from the center of Mira B) as given by Wood et
al.\ (2002) for the strong Mira B wind state, i.e., $R_s=3.7
\AU$ (this value will be lower for the weak Mira B wind
state).  Thus, the wind from Mira A is more likely to be
accreted by Mira B than to collide with the wind blown by
Mira B. While this conclusion leads us to propose that the
wind from Mira B is a bipolar outflow (perhaps in the form
of a collimated fast wind, of the type proposed by Soker \&
Rappaport 2000), it also casts further doubt on colliding
winds as the origin for the X-ray emission.  In addition,
the apparent X-ray flaring detected here (\S 3.2) is more
consistent with some form of magnetic and/or accretion
activity than with wind shocks.

A second possibility is
that the X-rays from the system originate with magnetic
activity on Mira A (Paper I). In light of our
recent XMM-Newton nondetections of X-ray emission from the
(apparently single) Mira variables TX Cam and T Cas (Kastner
\& Soker 2004), such an
explanation appears dubious. In addition, the $\sim2$
ks timescale of variation in the X-ray emission from the
Mira system is much shorter than the characteristic
dynamical timescales of AGB stars. 

In the remaining discussion, therefore, we only consider
processes in which the X-ray emission originates with Mira B or
its immediate environment, as is well established in the
case of the UV emission from the system (Karovska et al.\
1997; Wood et al.\ 2001). {\it Chandra} observations of
Mira AB at high spatial resolution will provide a crucial
test of this hypothesis. 

\subsection{The X-ray source: implications for the nature of Mira B}

The basic physics of accretion of the Mira A wind by Mira B
was discussed by JH84. Those authors convincingly argued
that the accreting star, Mira B, must be a main sequence
star, with a probable mass of $M_B \simeq 0.5 M_\odot$, and
radius $R_B \simeq 0.6 R_\odot$. The JH84 argument was based
in large part on the relatively meager X-ray luminosity of the Mira
system; from {\it Einstein} observations, JH84 derive an
X-ray luminosity $L_x \simeq 3 \times 10^{29} \erg
\s^{-1}$ ($0.15-2.5~$keV), where we
scale their $L_x$ result for a distance $D=128$~pc. This is
several orders of magnitude
lower than the $L_x$ expected from the Bondi-Hoyle model of
mass accretion onto
a white dwarf (WD) companion, given the mass loss rate and
wind speed of Mira (for which 
JH84 assumed $\dot M_A = 4 \times 10^{-7} M_\odot
\yr^{-1}$ and $v=5$ km s$^{-1}$, respectively, both of which are
consistent with more recent observational results; 
Knapp et al.\ 1998; Ryde \& Sch\"oier 2001). 

The results presented in \S 3.1 confirm the early
JH84 results for the $L_X$ of the Mira system. In particular,
although the XMM-Newton results demonstrate that the Mira
X-ray source is variable (\S 3.2), the EPIC data also
indicate that there is no appreciable hard ($E > 2.5$ keV)
X-ray emission from the system. 
Since the $L_X$ we and JH84 derive, $3-5 \times 10^{29} \erg
\s^{-1}$, is 1-3 orders of magnitude smaller than the X-ray
luminosities typical of accreting WDs in binary (cataclysmic
variable) systems (e.g., Pandel et al.\ 2003; Ramsay et al.\
2004) and the X-ray spectrum of Mira AB evidently
lacks the high-temperature ($kT_X > 2$ keV) component
characteristic of such systems, our results
strongly support the contention of JH84 that Mira B is
exceedingly unlikely to be a WD and is, instead, a low-mass,
main sequence star. This conclusion, in turn, has important implications
for the nature of the point-like X-ray sources within
planetary nebulae. Indeed, such X-ray sources
may be very similar in nature to the Mira AB binary and,
therefore, also could be powered via one of two
alternative, accretion-related mechanisms, as we now describe.

\subsection{X-rays derived from accretion onto Mira B}

Although Mira B is probably not a WD, the X-rays from the
Mira system might still be generated through accretion of AGB
wind material onto Mira B. The accretion process can produce
the X-rays directly, via star-disk interactions, or indirectly,
via the spin-up and resulting enhanced magnetic activity of
Mira B. Inserting reasonable parameters into equation (4) of
SK02, we indeed find that
an accretion disk is likely to be formed around a main
sequence companion to Mira A, such that either mechanism is
viable. We now discuss each process, in turn.

\subsubsection{The direct process: magnetospheric accretion}

If Mira B is a late-type star with a magnetically active,
convective envelope (see below), then the process of
accretion of wind material from Mira A may closely resemble
that of magnetospheric accretion onto low-mass, pre-main
sequence stars. For such (classical T Tauri) stars, it is
generally thought that material flows from accretion disk to
star along magnetic field lines (or ``funnels'') that are
rooted to the stellar surface at high latitudes
(e.g., Hayashi et al.\ 1996; Matt et al. 2002; Kastner et
al.\ 2004; and references therein). In the case
of the Mira system, a magnetospheric origin for X-ray
emission via star-disk interactions is favored by the
temperature derived from X-ray spectral fitting. This
temperature ($T_X \sim 10^7$ K) is somewhat high to 
be due to accretion shocks, given the likely free fall
velocity of matter onto the surface of a late-type main
sequence star (see, e.g., Appendix A of JH84).

The abundance anomalies suggested by the model fits (\S 3.1)
also point to Mira A's AGB wind material as a likely source of the
X-ray-emitting gas. Specifically, the enhanced abundances of
O and (possibly) Ne are consistent with the origin of this gas in
nuclear processed material dredged up from AGB interior
layers (Marigo et al.\ 1996; Herwig 2004). It is unclear that Mira A is
sufficiently massive and/or evolved to have generated 
excess O and Ne in thermal pulses, however; its
330 d period (Kukarkin et al.\ 1971) suggests a progenitor mass in the range
1.0--1.2 $M_\odot$ (Jura \& Kleinmann 1992), whereas
substantial O and Ne production likely requires a
progenitor of mass $>2$ $M_\odot$ (Marigo et al.). 

Another argument in support of the origin of the
X-ray emission in accretion processes is the result $N_H
> 2\times10^{21}$ cm$^{-2}$, derived from model
fitting of both ROSAT and XMM data (Paper I and \S
3.1). This X-ray-derived $N_H$ is a factor $>20$ larger 
than the neutral H absorbing column determined from analysis of the H I
Ly $\alpha$ line (Wood 
et al.\ 2002). This discrepancy suggests that the UV and
X-ray emission arise in different zones around Mira B and,
specifically, that the X-ray-emitting region may be embedded
within accretion streams that effectively attenuate the
X-rays, as described by JH84.

If the X-rays are indeed generated directly by accretion
onto Mira B,
then the similarity of the X-ray fluxes as measured in 1993
(by ROSAT) and in 2003 (by XMM) would suggest that the rate
of accretion has recovered from the relatively
low levels measured (via UV observations) in 1999-2001 (see Wood \& Karovska
2004 and references therein).

\subsubsection{The indirect process: spin-up of Mira B}

Alternatively, the X-ray emission from the Mira system may
be derived from the spin-up --- and resulting increase in
magnetic activity --- of Mira B, caused by accretion of mass
and angular momentum from the AGB wind of Mira A. Such a
process has been described by Jeffries, Burleigh, \& Robb
(1996), Jeffries \& Stevens (1996) and SK02. In SK02 (\S
2.1.1), we conclude that to generate an X-ray luminosity of
$L_x \simeq 5 \times 10^{29} \erg \s^{-1}$ via magnetic
activity, a main sequence star of spectral type M4 to F7
(i.e., in the mass range $0.3 M_\odot \lesssim M_B \lesssim
1.3 M_\odot$) has to be spun up to a period of $P \lesssim
3~$days, corresponding to an equatorial rotation speed of
$v_{\rm rot} \gtrsim 15-20 \km \s^{-1}$. Because we expect
an accretion disk to be formed around Mira B, we can use
equation (6) of SK02 to estimate the rotation velocity of
Mira B. That equation neglects the initial angular momentum
of the accreting main sequence star, and assumes that the
entire angular momentum of the star comes from mass accreted
from an accretion disk, with a Bondi-Hoyle mass accretion
rate (including initial angular momentum will increase the
rotation rate). Scaling with the parameters mentioned above,
and an orbital separation $a=100 \AU$ (somewhat larger
than the projected separation of $70 \AU$; Karovska et al.\
1997), we find
\begin{equation}
v_{\rm rot} \simeq 20 \frac {\Delta M_{\rm AGB}}{0.3 M_\odot}
\left( \frac {M_B}{0.5 M_\odot} \right)^{3/2} \left( \frac
{R_B}{0.6 R_\odot} \right)^{-1/2} \left( \frac {a}{100 \AU}
\right)^{-2} \left( \frac {v_r}{7 \km \s^{-1}} \right)^{-4} \km
\s^{-1},
\end{equation}
where $\Delta M_{\rm AGB}$ is the total mass lost by Mira A during
its AGB phase (when its wind speed remains low). Hence, Mira
B could have accreted sufficient angular momentum to spin fast enough
to be magnetically active at levels sufficient to explain
the observed $L_x \sim 5 \times 10^{29} \erg \s^{-1}$. The
lack of significant emission at energies $E > 3$ keV would make
Mira B somewhat unusual among highly active coronal sources, however
(e.g., Huenemoerder et al.\ 2003).

\acknowledgements{ We acknowledge support for this research
provided by NASA/GSFC XMM-Newton General Observer grant
NAG5--13158 to RIT. N.S. acknowledges support from the Israel
Science Foundation. We thank Falk Herwig and
Norbert Schulz for useful comments and suggestions.}

\begin{figure}[htb]
\includegraphics[scale=1.,angle=0]{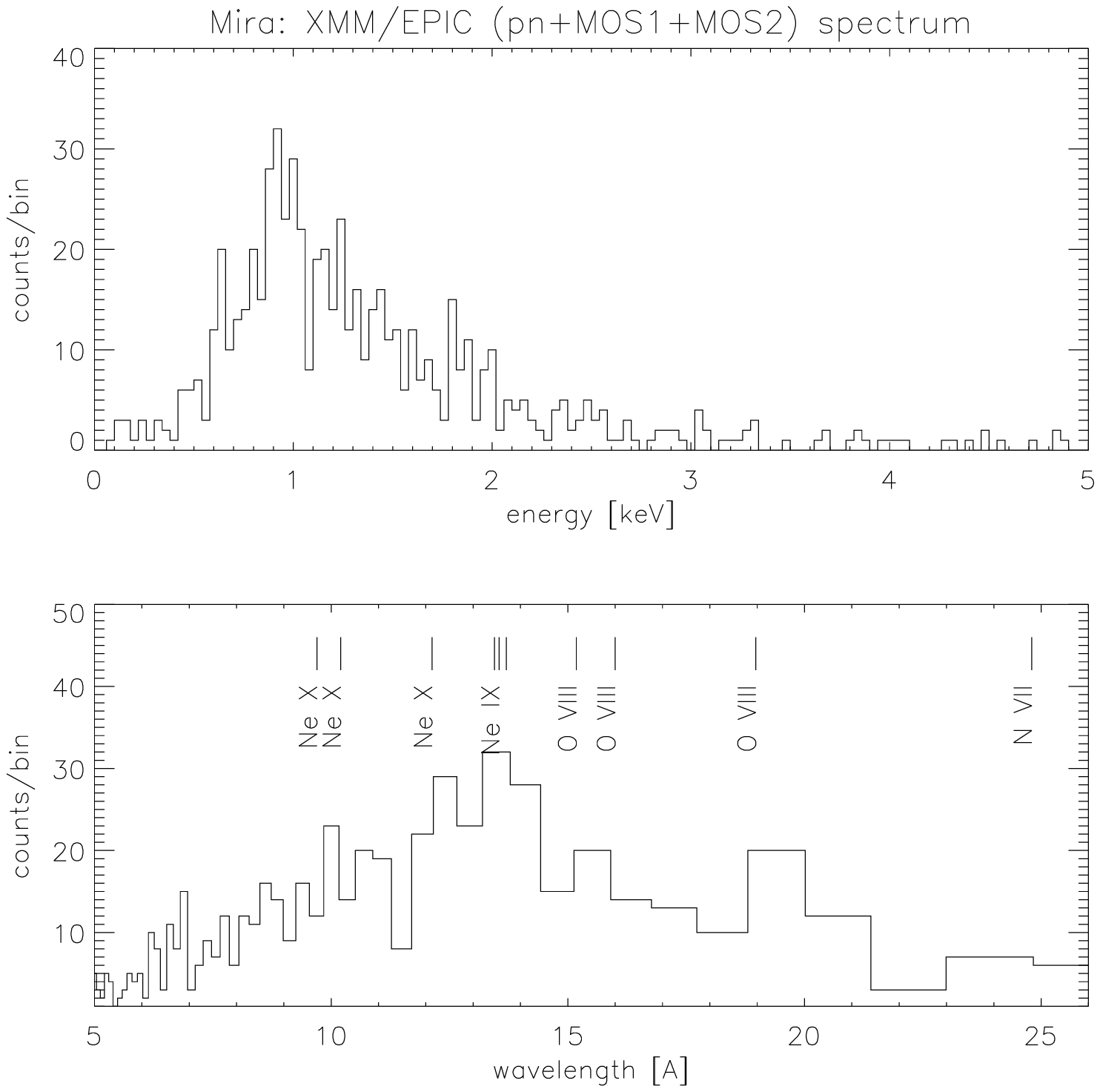} 
\caption{Combined EPIC (MOS1, MOS 2, and pn) counts spectrum of the
Mira system, plotted in 40 eV bins, as a function of energy
(top) and wavelength (bottom). Positions of prominent lines of highly
ionized N, O, and Ne are indicated in the bottom panel.} 
\end{figure}

\begin{figure}[htb]
\includegraphics[scale=.60,angle=-90]{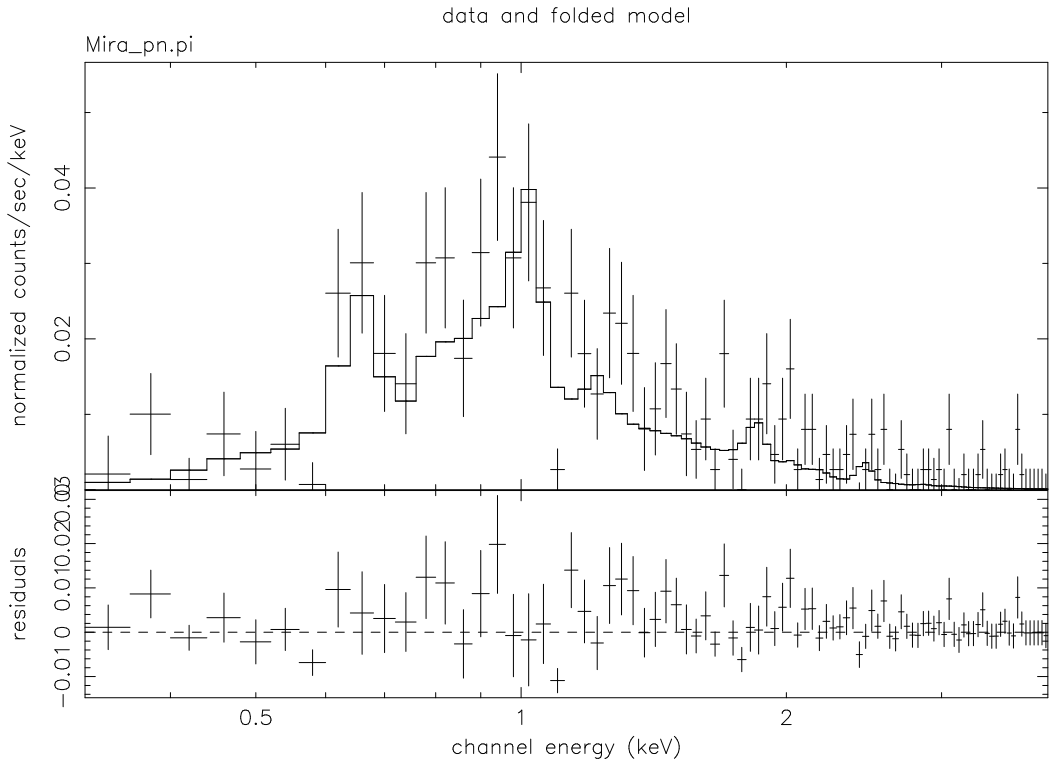} 
\caption{EPIC pn spectrum of the Mira system
  (crosses), with best-fit VMEKAL model overlaid (histogram). This best-fit
  model is obtained for $N_H = 4.5\times10^{21}$
  cm$^{-2}$ and $kT_X = 0.83$ keV, with highly elevated
  O abundance (see text). }
\end{figure}

\begin{figure}[htb]
\includegraphics[scale=1,angle=0]{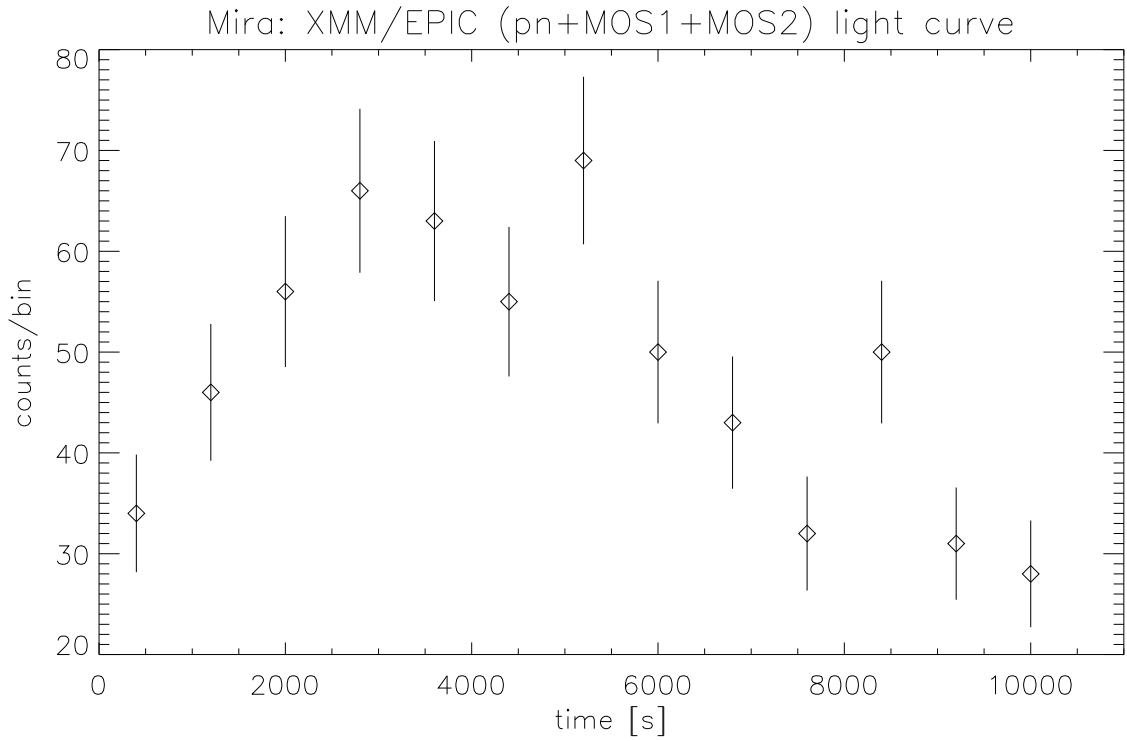} 
\caption{Combined EPIC (MOS1, MOS 2, and pn) light curve of
  the Mira system. Time bins are 800 s and the time interval
  is confined to the 10.4 ks period of active integration
  by all three instruments.} 
\end{figure}

\end{document}